\documentstyle[prl,aps,epsf,multicol]{revtex}    
\begin{document}
\draft
\title{Optical absorption edge in one-dimensional conductors}
\author{J. M. P. Carmelo$^{1}$, P. D. Sacramento$^{2}$, 
N. M. R. Peres $^{1}$, and D. Baeriswyl$^{3}$}
\address{
$^{1}$ Department of Physics, University of \'Evora,
Apartado 94, P-7001 \'Evora Codex, Portugal}
\address{
$^{2}$ Departamento de F\'{\i}sica and CFIF, IST,
Av. Rovisco Pais, P-1096 Lisboa Codex, Portugal}

\address{
$^{3}$ Institut de Physique Th\'eorique, Universit\'e
de Fribourg, P\'erolles, CH-1700 Fribourg,
Switzerland}
\date{1 May 1998}
\maketitle

\begin{abstract}
The frequency-dependent conductivity is studied for both
the one-dimensional Hubbard model and a model of spinless fermions,
using a selection rule, the Bethe ansatz energy eigenstates, and
conformal invariance. For densities where the system is metallic
the absorption spectrum has two contributions, a Drude peak at 
$\omega = 0$ separated by a pseudo-gap from a broad absorption band
whose lower edge is characterized by a non-classical critical 
exponent. Our findings are expected to shed new light
on the ``far infrared puzzle'' of metallic organic chain compounds.
\end{abstract}
\begin{multicols}{2}

Quasi-one-dimensional materials with partially filled electronic 
bands, i.e. ``metals'' from a naive point of view, usually undergo a 
transition to a low-temperature phase with a gap in the charge 
excitation spectrum. These transitions are driven by
electron-electron and/or electron-phonon interactions. Depending 
on the coupling constants and the band fillings the broken symmetry 
ground states are of different kinds: spin- or charge-density waves, 
bond alternation, superconductivity. Simple mean-field theory is 
able to describe these transitions, at least qualitatively. The 
broken symmetry induces a gap at the Fermi energy which may be
displayed in the optical absorption spectrum. In the case of 
charge-density waves accompanied by a lattice distortion,
optical gaps have indeed been clearly observed, and beautiful 
examples have been discovered among organic molecular 
chain compounds \cite{Jacobsen}, conjugated polymers \cite{Kiess}, 
and inorganic chain-like materials \cite{Gruner}. 
In the case of a moving spin-density wave, the sliding 
collective mode removes all the finite-frequency
transitions and produces a single peak at $\omega = 0$ 
\cite{Fenton}. Pinning by impurities not only shifts 
this peak to a finite frequency but also restores optical 
transitions above the gap \cite{Maki}.

The mean-field scenario makes sense at very low temperatures, where 
interchain coupling stabilizes three-dimensional long-range order.
In contrast, above the transition temperature the system may
enter an essentially one-dimensional regime, which requires a more
refined treatment of interactions. The real part of 
the optical conductivity can be written as $\sigma_1 (\omega) 
= 2\pi D\delta (\omega)+\sigma_1^{reg} (\omega)$, where
the frequency-dependent conductivity
$\sigma_1^{reg} (\omega)$ is defined in Eq. (\ref{cond}) below.
Giamarchi and Millis have used the bosonization technique 
and found $\sigma_1^{reg} (\omega) \sim \omega ^3$ in the 
limit $\omega \to 0$ for incommensurate band fillings \cite{Gia}. 
Unfortunately, it is at present not clear to what extent their 
treatment is a conserving approximation and, therefore, it is 
important to compare their results with those obtained using other 
techniques. Several authors have used exact diagonalization for the 
one-dimensional Hubbard model. In addition to finding no low-frequency 
absorption at half-filling, where the system is an insulator, these authors
have found practically no low-frequency ($\omega \neq 0$) absorption
away from half-filling, where a Drude peak at 
$\omega = 0$ confirms that the system is metallic \cite{Horsch}. 
This is not only in contrast to the bosonization results, but also to 
the two-dimensional case where doping away from half filling 
induces mid-infrared absorption \cite{Horsch,Dagotto}.
Analytical calculations in the large $U$ limit of the one-dimensional
Hubbard model have so far been restricted to the electronic density
$n = 1$ \cite{Lyo,Gebhard}. In this limit the optical gap occurs at 
$E_{opt} = U-4t$, followed by an absorption band extending 
up to $U+4t$ \cite{Lyo,Gebhard,Campbell,Mazumdar}. 

In this paper the optical absorption of the one-dimensional Hubbard
model is discussed in the framework of the Bethe ansatz,
which has been used previously by Schulz \cite{Schulz} to calculate 
$D$. Together with the optical 
sum rule, this allowed him to derive the total intensity of 
finite-frequency transitions. This intensity is 
generally small, except close to half filling for large enough 
(but not too large) values of $U$.
We have found that these finite-frequency transitions are
likely to be limited to a well-defined band above an effective optical gap,
which
is actually {\it smallest} at half filling and have also been able
to derive the exponent for the frequency-dependence above the absorption edge.

The 1D Hubbard model can be written as $\hat{H}=\hat{T}+U\hat{D}$, where 
$\hat{T}=-t\sum_{j,\sigma}[c_{j\sigma}^{\dag}c_{j+1\sigma} + h. c.]$
is the ``kinetic energy'',
$\hat{D} = \sum_{j}\hat{n}_{j,\uparrow}\hat{n}_{j,\downarrow}$
measures the number of doubly occupied sites,
$c_{j\sigma}^{\dagger}$ and $c_{j\sigma}$ are electron 
operators of spin projection $\sigma $ at site $j=1,...,N_a$, 
$\hat{n}_{j,\sigma}=c_{j\sigma }^{\dagger }c_{j\sigma }$,  
$t$ is the transfer integral, and $U$ is the on-site Coulomb 
interaction. We choose a density $n=N/N_a$ in the 
interval $0\leq n\leq 1$ with even $N$ and zero magnetization.
The Fermi momentum is given by
$k_F=\pi n/2$. We use units such that $-e=\hbar=1$, where 
$-e$ is the electronic charge. The regular part of the optical
conductivity can be written as

\begin{equation}
\sigma_1^{reg} (\omega) = {\pi\over N_a}
\sum_{\nu\neq 0}{\vert\langle\nu\vert \hat{J}
\vert 0\rangle\vert^2
\over \omega_{\nu,0}}\delta (\omega -\omega_{\nu,0}) \, .
\label{cond}
\end{equation}
Here $\hat{J} = -it\sum_{j,\sigma}\left
[c_{j\sigma}^{\dag }c_{j+1\sigma}-c_{j+1\sigma}^{\dag }
c_{j\sigma}\right]$ is the current operator, the  
summation runs over energy eigenstates, and 
$\omega_{\nu,0}=E_{\nu}-E_0$ is the excitation energy 
above the ground state $\vert 0\rangle$. 
The formula (\ref{cond}) applies if the ground state is 
non-degenerate. This is true for an even number of 
sites $N_a$ if periodic (anti-periodic) boundary 
conditions are used for odd (even) values of ${N\over 2}$, 
respectively. The ground state is then necessarily an 
eigenstate of the parity operator, $\hat{P}_{\pi}$, which 
moves electrons from sites $j$ to $N_a+1-j$, 
$j=1,...,N_a$, and has eigenvalues $\pm 1$. 
Importantly, in the present model $\hat{P}_{\pi}$ 
commutes with the Hamiltonian and anticommutes with the 
current operator. This implies immediately that 
the states $\vert 0\rangle$ and $\hat{J}\vert 0\rangle$ 
have opposite parities, a key selection rule for 
optical transitions. 
The final states $\vert\nu\rangle$ can be characterized in
terms of holons, antiholons, spinons \cite{Essler,Carmelo96}
and a charge-transfer band \cite{Carmelo97}. The holon and 
spinon bands describe low-energy charge and spin
excitations, whereas the charge transfer band is associated
with the upper Hubbard band. We use the labels $\alpha =c,s,t$ 
for the holon/antiholon, the spinon, and the charge-transfer 
bands, respectively \cite{Carmelo97}, and the quantum number 
$\beta$ for distinguishing between holons 
($\beta =-{1\over 2}$) and antiholons ($\beta =+{1\over 2}$) 
\cite{Carmelo96,Carmelo97}. 
In the present context, the $s$ band is empty of 
spinons, the $c$ band can be populated by holons and by zero 
or one antiholon, and the $t$ band can have occupancy zero or 
one. The momentum variable of the different bands has the 
form $q_j={2\pi\over N_a}I^{\alpha}_j$, where 
$I^{\alpha}_j$ are successive integers or half-odd integers. 
In contrast to the case of electron bands in a periodic
solid, the number of available momenta 
$N^*_{\alpha}$ can change with band occupancies. These 
numbers are $N^*_c=N_a$, $N^*_s=N/2-N_t$, and
$N^*_t=N_a-N+N_t$, where $N_t$ is the number of occupied
momenta in the charge-transfer band. Therefore, the momentum
band widths are $\Delta q_c=2\pi$, 
$\Delta q_s=2k_F-2\pi N_t/N_a$,
and $\Delta q_t=2\pi -4k_F+2\pi N_t/N_a$.
The numbers $N^h_{c,-{1\over 2}}$ (holons),  
$N^h_{c,+{1\over 2}}$ (antiholons), and $N_t$ are good
quantum numbers which obey the sum rules \cite{Carmelo97}, 
$N_a-N=-2\sum_{\beta=\pm {1\over 2}}\beta N^h_{c,\beta}
=N^h_c-2N_t$, where $N^h_c=\sum_{\beta=\pm {1\over 2}} 
N^h_{c,\beta}$. The ground state is characterized by
$N^h_{c,+{1\over 2}}=N_t=0$ and $N^h_{c,-{1\over 2}}=N_a-N$,
with a symmetrical holon occupancy in the $c$ band
for momenta $2k_F<\vert q\vert <\pi$.
The energy bands $\epsilon^0_{\alpha}(q)$ can
be extracted from the Bethe-ansatz solution \cite{Carmelo97}. 
Interestingly, we find that within the parameter region of 
appreciable oscillator strength for optical transitions 
\cite{Schulz}, the charge transfer band width 
$W_t = \epsilon^0_t(0) - \epsilon^0_t(\pi-2k_F)=\epsilon^0_t(0)$ 
is very small.

The main transitions contributing to $\sigma_1^{reg} (\omega)$
are of two types: (a) those
leaving the band fillings $N^h_{c,-{1\over 2}}$, 
$N^h_{c,+{1\over 2}}$, and $N_t$ unchanged and
(b) those changing these numbers by
$\Delta N^h_{c,-{1\over 2}}=\Delta N^h_{c,+{1\over 2}}= 
\Delta N_t=1$. The former start, in principle, at
$\omega=0$, while the latter have an onset at
$E_{opt}=W_t-2\epsilon^0_c(2k_F)>0$. 
We show first that, as a consequence of the selection rule
presented above, the transitions of type (a) are likely
to have a very small weight. These are excitations within the
holon band which can be characterized in terms of single and 
multiple electron-hole excitations \cite{Carmelo96}.
Single electron-hole excitations give no 
contributions to the finite-frequency absorption, while 
multiple excitations are expected to decrease very 
rapidly in intensity. Double excitations, which 
would give a contribution 
$\sigma_1(\omega ) \sim \omega^3$ \cite{Gia}, are 
forbidden by the parity selection rule, since these 
transitions require symmetrical changes of holon occupancies 
at $q$ and $-q$, leaving the parity unchanged.
In addition, there may be higher order low-frequency Umklapp
processes \cite{Gia}, but since these require multiple
excitations in the holon band, they are expected to have
extremely low spectral weight. These considerations explain
why numerical studies yield so weak features at low frequencies
\cite{Horsch,Horsch93}.

The weakness of type (a) transitions implies that the onset of 
type (b) transitions at $E_{opt}$ represents a pseudo-gap 
which may look like a true gap in an actual experiment.
$E_{opt}$, shown in Fig. 1, increases with
increasing $U$ and decreases with increasing density. For
$n=1$ it coincides with the Mott-Hubbard gap $E_{MH}$ 
\cite{Carmelo97}, which can be expressed as $E_{MH}=-2\epsilon^0_c(\pi)$
and is an increasing function of $U$. In general,
the type (b) transitions are also weak, except for $n$ close
to 1 and large $U$ \cite{Schulz}, corresponding to
a region of parameters where the bandwidth $W_t$ is very small. 
Therefore, we have limited our studies to this parameter
region. The important processes are those where in addition to a $t$
particle a holon-antiholon pair is created in the $c$ band.
This generates an absorption band with a lower edge
at $\omega =E_{opt}$ and a width 
$\Delta\omega_{opt} = -2[\epsilon^0_c(0)-\epsilon^0_c(2k_F)]
= 8t - 2\epsilon^h_F$. [Here
$\epsilon^h_F=\epsilon^0_c(\pi)-\epsilon^0_c(2k_F)$,
changes between $\epsilon^h_F =4t$ for $n\rightarrow 0$
and $\epsilon^h_F =0$ for $n\rightarrow 1$.]
The use of the quasiparticle/electron 
representation of Ref. \cite{Carmelo96} reveals that these 
transitions are associated with one zero-momentum 
electron - hole process. Moreover, creating a $t$ 
particle means enhancing double occupancy by 1, at least
in the large $U$ limit. Since the current operator cannot
produce more than one doubly occupied site, there will be
almost no spectral weight above the upper edge at 
$\omega = E_{MH}+8t+W_t\approx E_{MH}+8t$, illustrated by the 
dashed line in Fig.1. 

The spectral range of optical absorption deduced above on
the basis of the holon and charge transfer excitation
spectra fully agrees with large $U$ expansions 
\cite{Gebhard} and numerical calculations
\cite{Horsch,Campbell}. A much more difficult problem is the 
spectral line shape. Here, we limit ourselves to the region
immediately above threshold, using the hypothesis of
conformal invariance. In this region the transition energies
have a finite-size dependence 
$E_{\nu}-E_{0}-E_{opt}={2\pi\over N_a}
\sum_{\alpha,\iota=\pm 1}v_{\alpha}
\Delta^{\iota}_{\alpha}+O(1/N_a)$
where $\Delta^{\iota}_{\alpha}$
are $U$ and $n$ dependent positive parameters. 
The momentum is the same as in the ground state, i.e.
$P_{\nu}={2\pi\over N_a}\sum_{\alpha,\iota=\pm 1}\iota
\Delta^{\iota}_{\alpha}=0$. 
By analogy to the case of the two-velocity spectrum of Ref.
\cite{Frahm}, we interpret this as a low-energy 
$(\omega-E_{opt})$ three-velocity conformal-field theory, whose 
vacuum is $\vert 0\rangle$ with energy measured from 
$E_0+E_{opt}$. We have verified that
the exponents for the conductivity do not depend on whether
they are calculated for $v_t\rightarrow 0$ or for $v_t=0$.
The physical fields are of the 
general form $\phi (x,t)e^{iE_{opt} t}$. The
anomalous dimensions of the $\phi (x,t)$ fields involve the 
parameters, $\xi^j_{\alpha\alpha'}=
\delta_{\alpha,\alpha'}
+ \sum_{l=\pm 1}l^j
\Phi_{\alpha\alpha'}
(q_{F\alpha},lq_{F\alpha'})$ (with $q_{Fs}=k_F$ and
$q_{Ft}=0$). 
Here $j=0,1$ and the two-particle forward-scattering phase shifts, 
$\Phi_{\alpha\alpha'}(q,q')$, can be extracted
from the Bethe-ansatz solution \cite{Carmelo97}. 
The critical behavior of the conductivity (\ref{cond}) can then
be calculated, with the result
$\sigma_1^{reg}(\omega)=C \Bigl(\omega -E_{opt}\Bigl)^{\zeta }$ 
at the onset and a critical exponent 

\begin{equation}
\zeta = - {3\over 2} + 
{1\over 2}\Bigl(2\xi^0_{cc}-\xi^0_{ct}\Bigl)^2 + 
{1\over 2}\Bigl(\xi^0_{st}\Bigl)^2 \, .
\label{exp}
\end{equation}
The exponent $\zeta$, shown in Fig. 2, approaches ${1\over 2}$ as 
$U\rightarrow\infty$, and for densities $n<1$ it increases with 
decreasing values of $U$. For $n\rightarrow 1$
we find $\zeta={1\over 2}$ and $W_t=0$ for all finite
values of $U$. Putting $v_c=0$ and $n=1$
in the $x$ and $t$ dependent current - current 
(or charge - charge) correlation function and taking
the Fourier transform, we again obtain 
$\zeta={1\over 2}$ at the threshold of optical
absorption. Therefore, we believe that at $n=1$
the expression $\sigma_1^{reg}(\omega)=
C \Bigl(\omega -E_{MH}\Bigl)^{{1\over 2}}$
is valid for all finite values of $U$. 
For the particular case of $n=1$ the same 
method allows us to evaluate the leading frequency dependence
close to the upper edge of the absorption band. 
We find $\sigma_1^{reg} (\omega)=
C\Bigl(E_{MH}+8t-\omega\Bigl)^{{1\over 2}}$. 
This symmetry between the lower and upper
absorption edges for $n=1$ was also found in Refs. 
\cite{Gebhard,Mazumdar}. 
Moreover, expanding the $n=1$ expression $(28)$ of Ref.
\cite{Gebhard} around the edge energies leads
precisely to the same $\omega $
power-law dependences with exponent ${1\over 2}$.
We notice that the constant
$C$ has to vanish for all densities 
as $U\rightarrow 0$ or $U\rightarrow\infty$, since
in these limits the Drude peak exhausts the 
conductivity sum rule \cite{Schulz}.

It is worthwhile to compare our predictions with
experiments on one-dimensional conductors. Unfortunately,
as a general rule these materials cannot be represented 
in terms of the Hubbard model alone, but additional
interactions have to be taken into account. An exception
seems to be the organic chain compound $(TTM-TTP)I_3$,
which displays the properties of the
one-dimensional Hubbard model at half filling with a 
relatively small value of $U$ ($U\approx 3.5 t$): an
optical gap of the expected size \cite{Tajima} and a
Pauli-like susceptibility \cite{Mori}. The optical
absorption is consistent with a smooth onset at
threshold, but it seems difficult to extract a critical
exponent from the data. In the past there has been considerable
discussion concerning the phenomenon of suppressed far
infrared conductivity observed in organic chain compounds,
despite their large dc conductivity \cite{Timusk}. Very
recently, the optical absorption spectrum of large
single crystals of Bechgaard salts has been measured over
a huge frequency range, confirming the far infrared problem
\cite{Degiorgi}.
A Drude peak corresponding to a very small carrier density
is separated by a gap of about 0.01 eV from an absorption 
band of a width of about 0.6 eV. The onset of absorption
above the gap is very sharp at low temperatures. In the
framework of the one-dimensional Hubbard model these results
can be understood if the system has a nearly half-filled band,
in which case the optical threshold is essentially given
by the Mott-Hubbard gap. The parameters $t \approx 0.25 eV$ and 
$U/t \approx 1.5$ allow to reproduce the gross features of 
the experimental spectra. 

While the simple Hubbard model may be sufficient for
reproducing the overall absorption spectrum of some
of the one-dimensional conductors, additional terms
such as the Coulomb interaction between nearest 
neighbor sites would have to be considered for a
more detailed description of the spectrum, for
instance for the excitonic enhancement at the 
absorption edge \cite{Mazumdar}. Since
the extended Hubbard model is not solvable by the 
Bethe ansatz, we have considered the model 
Hamiltonian $\hat{H} = -t\sum_{j}
[c_{j}^{\dag}c_{j+1} + h. c.]
+V\sum_{j}\hat{n}_{j}\hat{n}_{j+1}$,
where $c_{j}^{\dagger}$ creates a spinless fermion
at site $j=1,...,N_a$, 
$\hat{n}_{j}=c_{j}^{\dagger }c_{j}$,  
and $V$ is the nearest-neighbor Coulomb interaction. 
For a density $n={N\over N_a}={1\over 2}$
this model has a metal - insulator
transition for $V>2t$. As for the Hubbard model, in the region 
$V>2t$ we find almost no spectral weight 
in $\sigma_1(\omega )$ for $\omega <E_{opt}$, where 
$E_{opt}$ is a $V$-dependent optical gap.   
We again obtain a critical behavior $\sigma_1^{reg}(\omega)=
C \Bigl(\omega -E_{opt}\Bigl)^{\vartheta}$ 
for low values of $(\omega -E_{opt})$, with $\vartheta\rightarrow 
{1\over 2}$ both as $n\rightarrow {1\over 2}$ and $V>2t$ and 
for all densities as $V\rightarrow\infty$. For $n={1\over 2}$ 
and large $V$ the corresponding absorption extends from 
$\omega =V-4t$ to $\omega=V+4t$.

In summary, we have discussed the frequency-dependent conductivity of
the one-dimensional Hubbard model (and of a model of spinless fermions
with nearest neighbor interaction), using a parity selection rule and
known properties of the exact eigenstates. 
In the insulator phases the optical absorption has a square-root 
behavior at the onset. For $n \neq 1$ ($n \neq 1/2$
and $V>2t$) the spectrum consists of a Drude peak at $\omega = 0$ and
an absorption band above a pseudo-gap, which increases as the
system is doped away from the insulating state. 
The behavior at the onset of the finite-frequency transitions is 
described by a critical exponent that depends both on the interaction 
strength and the density. The intensity of the transition across the 
gap is appreciable only when the charge-transfer bandwidth is very 
narrow. This is reminiscent of the X-ray edge singularity which 
depends very sensitively on the recoil of the hole \cite{Nozieres}. 
Our results are expected to shed new light on the old problem of 
optical absorption in one-dimensional conductors.
 
We thank D.K. Campbell, P. Horsch, and J.M.B. Lopes dos Santos 
for illuminating discussions.


\narrowtext
\begin{figure}
\label{fig1}
\caption{The gap $E_{opt}$, (a) as function of 
$U$ for different values of $n$ and (b) as function of $n$ 
for different values of $U$.}
\end{figure}
\begin{figure}
\label{fig2}
\caption{The exponent $\zeta$, (a) as function of 
$U$ for different values of $n$ and (b) as function of $n$ 
for different values of $U$. The straight lines at $\zeta=1/2$ 
refer to the (a) $n\rightarrow 1$ and (b) $U\rightarrow\infty$ 
limits.}
\end{figure}
\end{multicols}
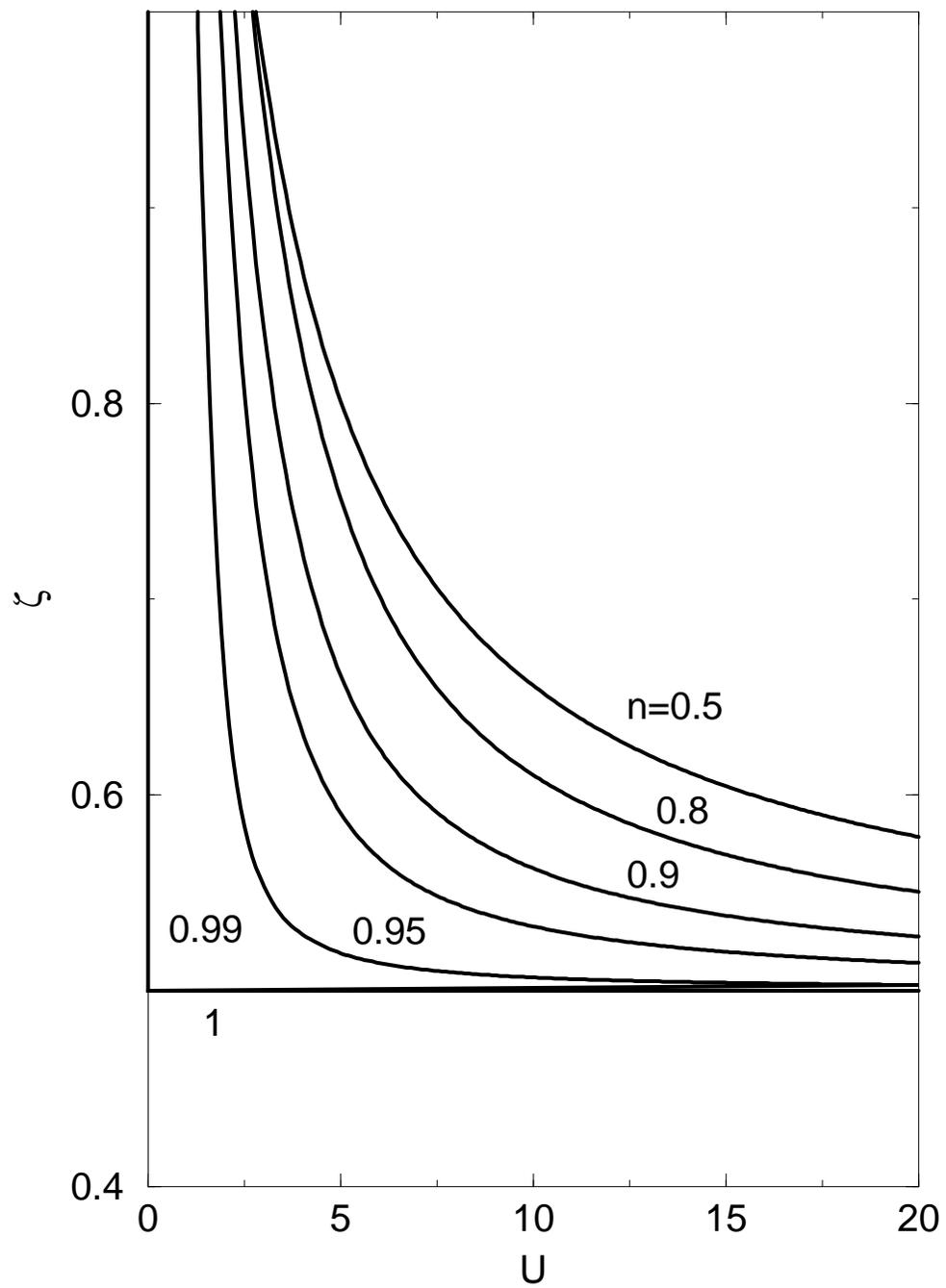
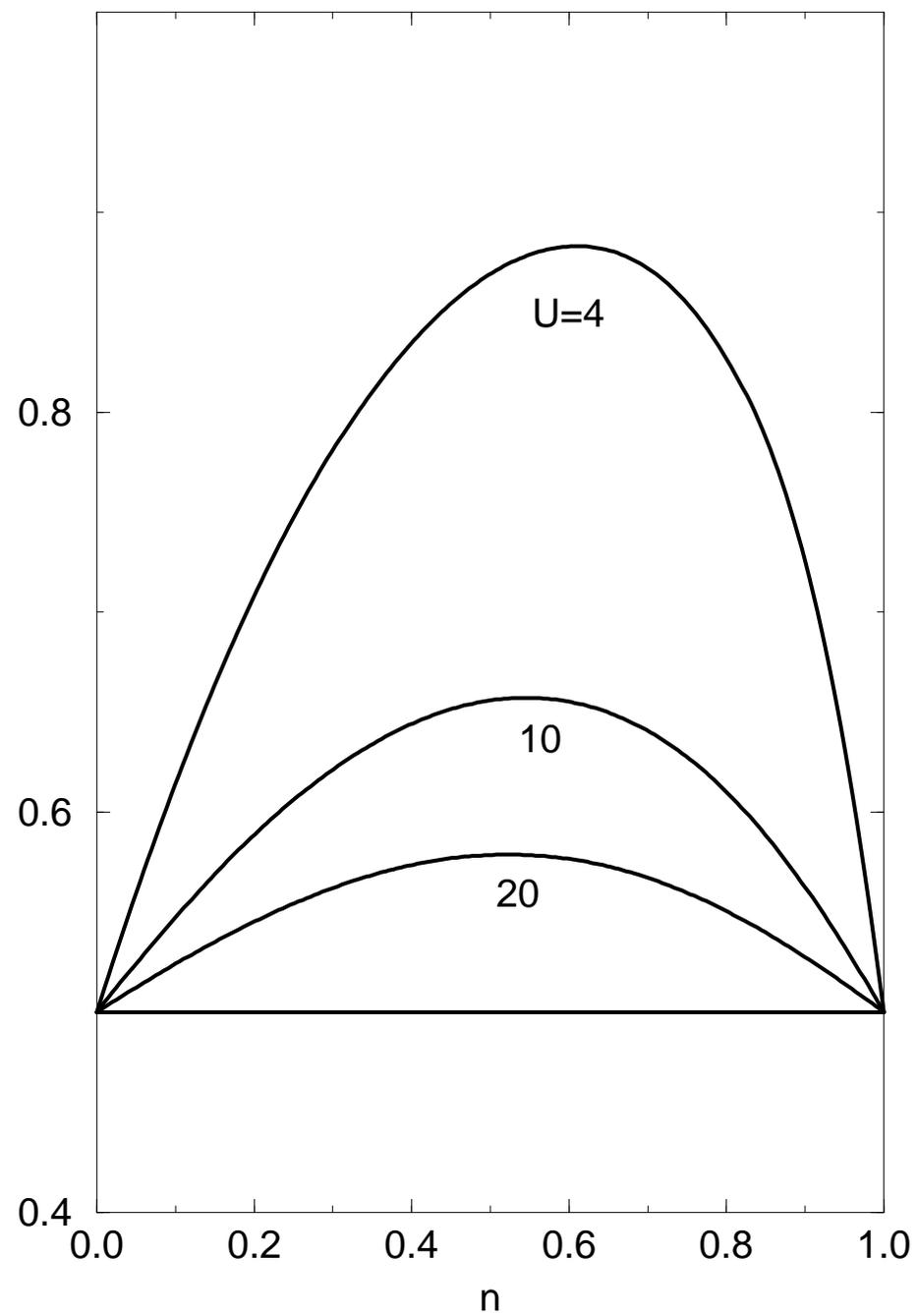
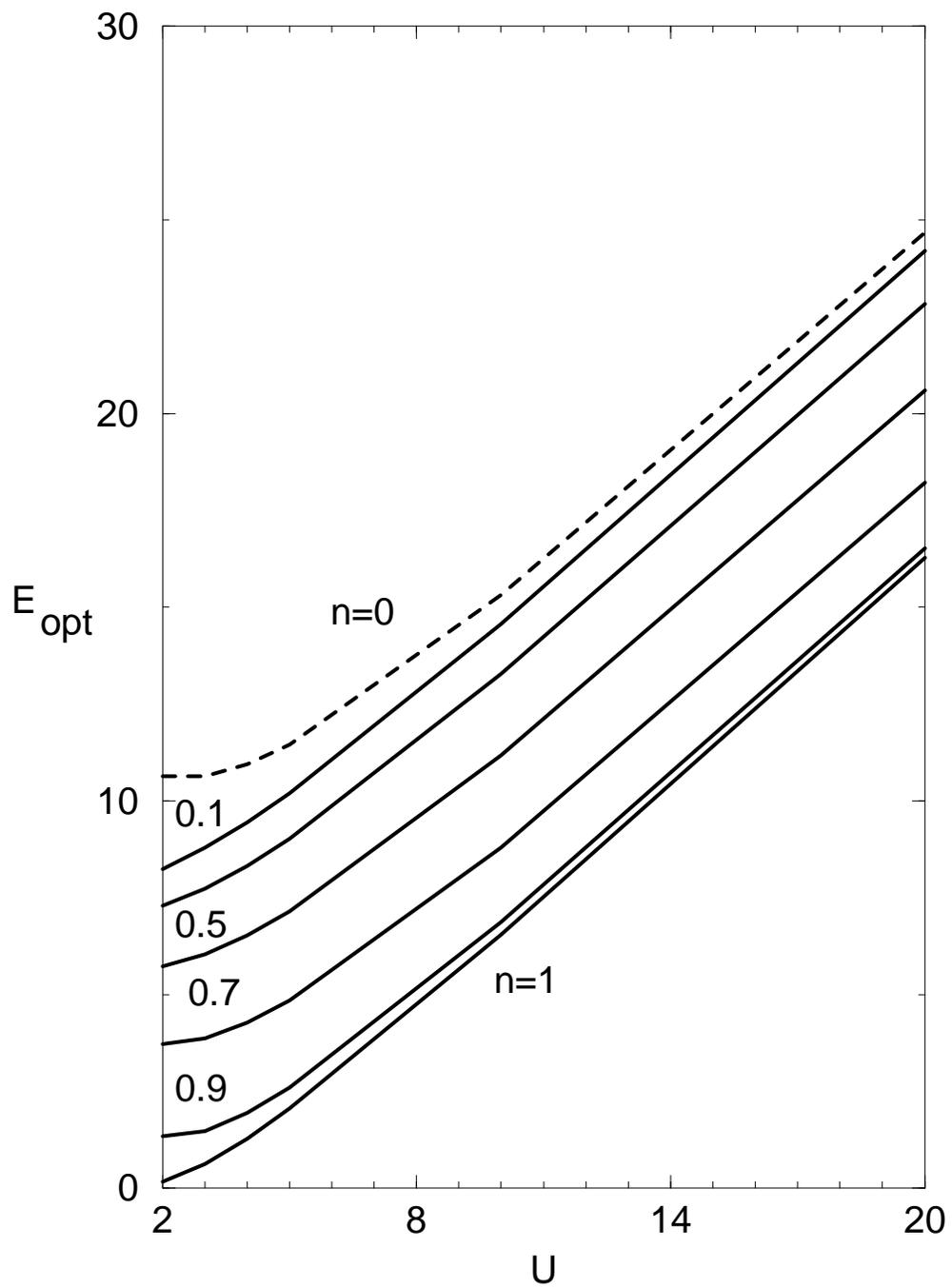
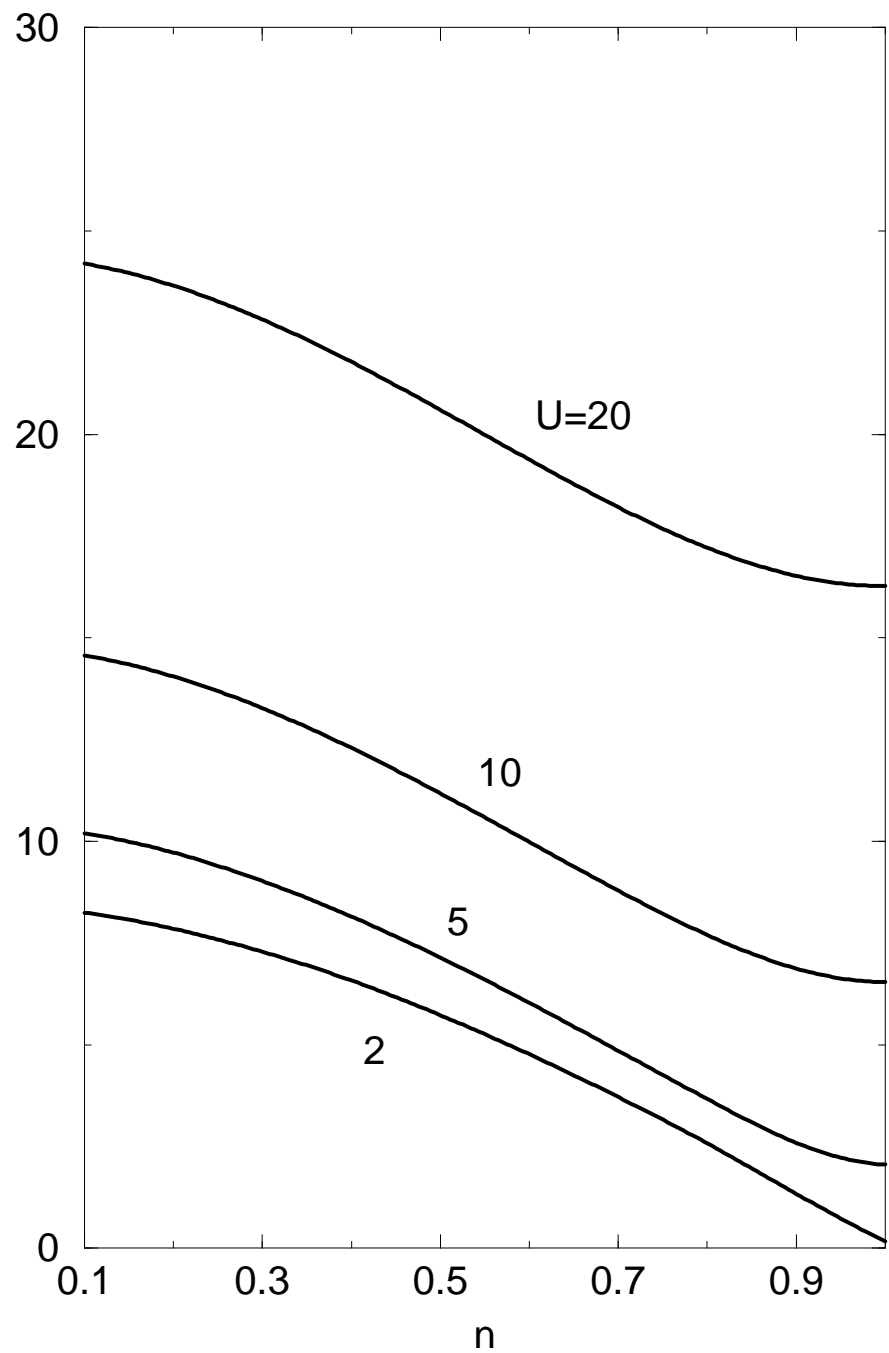
\end{document}